\documentstyle[epsfig]{article}           
\begin{document}

\newcommand{\Ne}{N\'{e}el}
\newcommand{\be}{\begin{equation}}
\newcommand{\ee}{\end{equation}}

\title{
The antiferromagnetic spin $1/2$ chain with competing dimers
and plaquettes: \\ Numerical versus exact results
} 

\author{
J Richter\dag, N.B.Ivanov\ddag, J.Schulenburg\dag \\
\dag\ Institut f\"ur Theoretische Physik, Universit\"at Magdeburg,\\
 P.O.Box 4120, D-39016 Magdeburg, Germany\\
\ddag\ Institute  for Solid  State Physics, Sofia, Bulgaria
}

\maketitle

\begin{abstract}
We examine the ground state and the excitations of an
one-dimensional Heisenberg
spin $1/2$ antiferromagnet with alternating dimers and four-spin plaquettes
(dimer-plaquette chain).
The properties of the
system depend on the competing dimer and plaquette bonds.
Several exact,
exact-numerical and perturbational results are presented.
We find that the system is gapped for all parameter values. The
spin pair correlation functions can be characterized by three different
correlation lengths for dimer -- dimer, dimer -- plaquette and plaquette --
plaquette correlations.
For the latter one we find an effective $S=1$ Haldane like behaviour in
the limit
of dominating dimer bonds.

Introducing frustration the system undergoes
a first order phase transition to a fully dimerized state.
Concerning the phase relationships of the ground-state wave function
the system represents an example for the exact validity of the
Marshall-Peierls sign rule   in a strongly frustrated antiferromagnet.

The model considered is related to the recently found
$1/5$-depleted
square-lattice Heisenberg system $CaV_4O_9$.
\end{abstract}


\section{Introduction}
 The exciting
collective magnetic properties of low-dimensional quantum spin systems
 have attracted much attention over the last decade. The search for systems
with spin-liquid ground states is one subject of continuous interest.
 Compressible (gapped) and incompressible (gapless) spin liquid
phases with more or less exotic
ground-state ordering were discussed in particular for
the frustrated $J_1-J_2$  model on the square lattice
(see e.g. \cite{qsl}).

The recent discovery of a spin gap in $S=1/2$ quasi-two-dimensional
$CaV_4O_9$  \cite{taniguchi95,ohama97}
has stimulated the investigation of quantum disorder and gap formation
of systems with
different types of antiferromagnetic nearest neighbour (NN) bonds
\cite{katoh95,ivanov96,albrecht96,troyer96,meshkov96,richter96,richter97}.
$CaV_4O_9$ has a  layered structure where the
magnetic $V^{4+}$ ions have spin 1/2 and form a $1/5$ depleted square lattice
\cite{struc73,starykh96}.
The minimal model for $CaV_4O_9$ is a $1/5$ depleted
Heisenberg model, i.e. a model with
4-spin plaquettes connected at there edges with one neighbouring plaquette.
Because of the distortion of the lattice \cite{struc73,starykh96}
the intra-plaquette $J_p$ and the inter-plaquette (dimer) $J_d$ bonds might be
different. Though in
a classical version of this nonfrustrated Heisenberg model the
\Ne\ state is the stable ground state for any $J_p>0$, $J_d > 0$, in the
quantum case  a competition arises between a local singlet formation of a
couple of spins along a dimer bond $J_d$ and a local singlet formation of
the four spins belonging to a plaquette and coupled by $J_p$.
However, the explanation of the
measured spin gap by competition between $J_p$ and $J_d$ would require
unreasonable large differences  between $J_p$ and $J_d$. As proposed in
several papers \cite{ueda96,starykh96,gelfand96,white96,sachdev96,bose97}
one needs additional frustration to get reasonable
values for the gap.

In this paper we extend our preliminary discussion \cite{richter96,richter97} 
of the competition between dimer and
plaquette bonds and the role of frustration in the one dimensional
counterpart of the depleted square lattice Heisenberg model.
This model is simpler than
the 2d model, but nevertheless it contains non-trivial physics.
To our knowledge presently there
is
no corresponding quasi-1d material but it seems to be possible
that it can be synthesised in future.

Though the considered Heisenberg model is a spin $1/2$ model it will
be shown below that the model contains also elements of the physics of the
spin $1$ chain which is currently also under intensive discussion (see
e.g. \cite{white93,golinelli94,mikeska,scholl} and references therein).

The paper  is organised as follows: In Section 2 we present the model and
elaborate some exact statements concerning eigenvalues and eigenstates of the
model.
In Section 3 we discuss exact numerical data for chains up to 32 sites as
well as analytic results using perturbation theory.
Conclusions are given in Section 4.

\section{Model and general results}

\begin{figure}
\begin{center}{\input{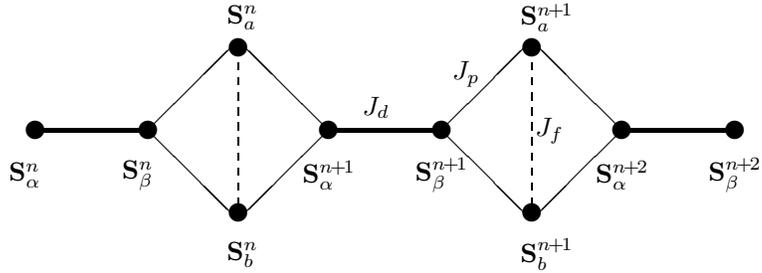}}\end{center}  
\caption{\label{fig1} (see text)}
\end{figure}

The spin $1/2$ Heisenberg chain considered here consists of $N$ spins
forming $N_p=N/4$ plaquettes which are connected by $N_d=N_p$ dimer
bonds (see Fig.\ref{fig1}). The two spins connected by the $nth$ dimer bond $J_d$
we call dimer spins and denote them by  ${\bf S}^n_{\alpha}$ and
${\bf S}^n_{\beta}$, where the index ${\alpha}$ (${\beta}$) stands for the
 left (right) spin.  The two spins sitting at the top
and the bottom of the $nth$ plaquette we call plaquette spins and denote them
by  ${\bf S}^n_{a}$ and
${\bf S}^n_{b}$, where the index ${a}$ (${b}$) stands for the
top (bottom) spin.
With these notations we write the Hamiltonian
\begin{equation}\label{ham1}
H_{d-p} = J_d \sum^{N_p}_{n=1}{{\bf S}^n_{\alpha}{\bf S}^n_{\beta}}
+  J_p \sum^{N_p}_{n=1}
{({\bf S}^n_{\beta}{\bf S}^n_{a}+{\bf S}^n_{\beta}{\bf S}^n_{b}+
{\bf S}^n_{a}{\bf S}^{n+1}_{\alpha}+{\bf S}^n_{b}{\bf S}^{n+1}_{\alpha})}
\end{equation}
($J_d,J_p > 0$). 
Frustration is introduced by a diagonal antiferromagnetic  bond $J_f$
connecting a top and bottom plaquette  spin (see Fig.\ref{fig1}). Then the  total
Hamiltonian reads
\begin{equation}\label{ham2}
H =  H_{d-p} + H_f =
H_{d-p} +
J_f \sum^{N_p}_{n=1}{{\bf S}^n_{a}{\bf S}^n_{b}} \hspace{6pt},
\hspace{6pt} J_f \ge 0.
\end{equation}

For the sake of convenience we consider chains with periodic boundary
conditions.

Approximately at the same time when we introduced \cite{richter96} 
the above defined dimer-plaquette chain Takano and coworkers \cite{takano96} 
considered a so-called diamond chain, built by plaquettes only. The ground
state problem of this diamond chain was recently analysed by Niggemann
 et al. \cite{nigge97}.
The main difference between both models consists in the existence of the
dimer bond in the dimer-plaquette chain. As a consequence, the both models
belong to different universality classes. 
We will briefly discuss some important differences between both models 
in section 4.


For the above described dimer-plaquette chain (\ref{ham1},\ref{ham2})
we can find the following general statements:

\subsection*{\bf(i) {\bf Classical ground state:}}

For $J_f < J_p$
the ground state is a \Ne\ state.
The
correlations are
$\langle {\bf S}^n_{a}{\bf S}^n_{b}\rangle=+S^2$
between a bottom and top spin of  the same plaquette ($J_f$ bond),
$\langle {\bf S}^n_{\alpha}{\bf S}^n_{\beta}\rangle=-S^2$
between two neighbouring
dimer spins ($J_d$ bond)
and $\langle {\bf S}^n_{\beta}{\bf S}^n_{a(b)}\rangle =
\langle {\bf S}^{n+1}_{\alpha}{\bf S}^n_{a(b)}\rangle =-S^2$
between a dimer spin and a neighbouring plaquette spin ($J_p$ bond).
For $J_f > J_p$
the ground state has twisted plaquette spins.
The corresponding
correlations of neighbouring spins are
$\langle {\bf S}^n_{a}{\bf S}^n_{b}\rangle=S^2(2J^2_p/J^2_f - 1)$,
$\langle {\bf S}^n_{\alpha}{\bf S}^n_{\beta}\rangle=-S^2$
and $\langle {\bf S}^n_{\beta}{\bf S}^n_{a(b)}\rangle =
\langle {\bf S}^{n+1}_{\alpha}{\bf S}^n_{a(b)}\rangle=-S^2 J_p/J_f$.

Now we turn to the quantum spin $1/2$ case.

\subsection*{\bf(ii)  {\bf Integrals of motion:}}

In addition to the usual integrals of motion (z-component and square of
total spin) there are $N_p$ local integrals of motion,
namely the square of the total
spin of the top and bottom spin of a plaquette $n$, i.e
\begin{equation}\label{com}
[H,({\bf S}^n_{ab})^2]_- = 0 \hspace{10pt}, \hspace{10pt}
{\bf S}^n_{ab}={\bf S}^n_{a} + {\bf S}^n_{b}.
\ee
Hence we can classify all eigenstates by the following
set of the quantum numbers:
energy $E$, z-component of the total spin $M$,
square of the total spin $S$, and $N_p$ local quantum numbers $S^n_p$ of
$({\bf S}^n_{ab})^2$,
where the values for $S^n_p$ are $0$ (singlet) or $1$ (triplet).
Concerning the
correlation function between a top and bottom spin
of plaquette $n$ we have
$\langle {\bf S}^n_{a}{\bf S}^n_{b}\rangle = -3/4$
\hspace{4pt}$(+1/4)$ \hspace{6pt} for \hspace{6pt} $S^n_p=0$
\hspace{4pt}($S^n_p=1$).

\subsection*{\bf
(iii) {\bf Lieb-Mattis theorem and ground state in the nonfrustrated
limit:}}

In the limit $J_f=0$, i.e. $H=H_{p-d}$, the lattice is bipartite and
the Lieb-Mattis theorem is valid \cite{lm62,klein82}, i.e. the
ground state is a singlet $S=0$ of the total spin. As a
consequence of the theorem we have
$\langle {\bf S}^n_{a}{\bf S}^n_{b}\rangle > 0$
since the top and bottom spins of a plaquette $n$ belong to the same
sublattice, i.e. the ground state is a singlet of the total spin but all
local quantum numbers are $S^n_p=1$ ($n=1,\ldots,N_p$).

We notice that the numerical results (see below) indicate
that the ground state is a singlet of the total spin for
finite frustration, too, which is in accordance with other
calculations of the ground state of various frustrated antiferromagnets.

\subsection*{\bf
(iv) {\bf Mapping onto a spin$-1/2-$spin$-1$ chain:}}

As a result of (iii) for zero temperature  the Hamiltonian (\ref{ham1})
can be exactly mapped on a chain with mixed spin $1/2$ and spin $1$
objects as shown in Fig.\ref{fig2}.

\begin{figure}
\begin{center}{\input{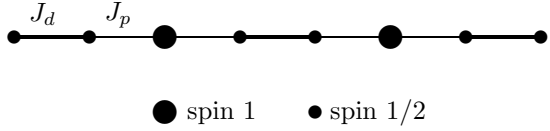}}\end{center}     
\caption{\label{fig2}
Spin$-1/2$ -- spin$-1$ chain, which is equivalent to the
nonfrustrated Hamiltonian (\ref{ham1}) for zero temperature.
}\end{figure}

This effective model describes not only the ground state, but also all other
eigenstates of
(\ref{ham1}) with $S^n_p=1$ for all $n=1,\ldots,N_p$.

Notice, that antiferromagnetic chains with alternating spins $S=1/2$ and
$S=1$ where recently discussed as quantum ferrimagnet
\cite{ferri1,ferri2,ferri3};
however, the
effective model here is of different kind since we have twice as much spins
$S=1/2$ as spins $S=1$.

\subsection*{\bf (v) {\bf Product eigenstates:}}

We consider now the class of eigenstates which do not correspond
to the spin$-1/2$ -- spin$-1$ chain, i.e. we consider states
where some of the local quantum numbers $S^n_p$
are zero. Suppose  $S^i_p =0$ in the plaquette $i$. Then
the top and bottom spins of plaquette $i$ form a singlet which is
decoupled from all other spins, i.e. we have
$\langle {\bf S}^n_{a(b)}{\bf S}^{i}_{a(b)}\rangle=0$
($n \ne i)$ and
$\langle {\bf S}^n_{\alpha(\beta)}{\bf S}^{i}_{a(b)}\rangle=0$.
Hence the eigenstate can be written in product form
\begin{equation}\label{psi}
|\Psi\rangle = |(a_{i},b_{i})\rangle|\Psi_{remainder}\rangle
\end{equation}
where $|(a_{i},b_{i})\rangle =
(\uparrow_{a_{i}}\downarrow_{b_{i}}-
\downarrow_{a_{i}}\uparrow_{b_{i}})/\sqrt{2}$
\hspace{4pt} is a pair singlet state of the top and
bottom spin of plaquette $i$ and
$|\Psi_{remainder}\rangle$ is a state describing all the remaining
$N-2$ spins forming a corresponding open chain with dimer ends.
Suppose $S^i_p =0$ in $L>1$ plaquettes $i$.
Then the eigenstate
separates in $L$ pair singlet states of the top and
bottom spin of plaquettes $i$ and eigenstates of the finite chain pieces
lying
between two plaquettes with  $S^i_p=0$.
The more plaquettes $i$ in a singlet state $S^i_p=0$ the shorter
the finite chain pieces between two plaquettes with
 $S^i_p=0$.
The extreme case is the
state with $S^n_p=0$ for all
$n=1,\ldots,N_p$ plaquettes, where the finite pieces between two plaquettes
are just the dimers themselves.
This state can be explicitly written as
\begin{equation}\label{psi1}
|\Psi_{0,\ldots,0}\rangle = \prod_{n=1}^{N_p}|(a_n,b_n)\rangle
\prod_{n=1}^{N_p}|(\alpha_n,\beta_n)\rangle
\end{equation}
where $|(\alpha_n,\beta_n)\rangle =
(\uparrow_{\alpha_n}\downarrow_{\beta_n}-
\downarrow_{\alpha_n}\uparrow_{\beta_n})/\sqrt{2}$
\hspace{4pt}  is a pair singlet state of a dimer bond
$n$.
The energy of this state is
\begin{equation}\label{ene1}
E_{0,\ldots,0}=-\frac{3}{4} J_d N_p -\frac{3}{4} J_f N_p.
\end{equation}

\subsection*{\bf
(vi) {\bf Eigenstates and energy levels --  frustrated versus
nonfrustrated model:}}

Between the eigenstates and the energy  of the nonfrustrated (\ref{ham1})
and the frustrated (\ref{ham2}) model there exist simple relations  due to the
fact that $H_{p-d}$ commutes with $H_f$. Hence the eigenfunctions of
$H_{p-d}$ are not changed including frustration
and for the energy contribution of the
frustrating part $H_f$ only the local quantum numbers $S^n_p$ are
important. Consider any eigenstate of $H_{p-d}$ with energy $E_{p-d}$
and $N^s_p$ plaquettes with quantum number $S^n_p=0$ and
$N^t_p$ plaquettes with quantum number $S^i_p=1$ ($N^t_p + N^s_p =N_p $).
Then the energy for the frustrated model $H=H_{p-d} + H_f$ is
\begin{equation}\label{ene2}
E_{p-d,f}=E_{p-d} + J_f(\frac{1}{4} N^t_p  - \frac{3}{4} N^s_p )
= E_{p-d} + J_f(\frac{1}{4} N_p  - N^s_p )
\end{equation}

\subsection*{\bf
(vii) {\bf Upper and lower bound for the critical $J^c_f$:}}

From Eq. (\ref{ene2}) it is obvious that $J_f$ favours
energetically the singlet formation of plaquette spins and for
large $J_f$ the singlet product state (\ref{psi1}) becomes the
ground state of $H$.

According to (iii) the ground state is the lowest eigenstate
with $S^n_p=1$ for all
$n=1,\ldots,N_p$ for $J_f=0$ and has the energy $E^{0}_{1,\ldots,1}$.
Following the ideas of Ref. \cite{nigge97} we used the linear 
programming scheme  
to prove that  at a critical value $J^c_f>0$ a first order
transition takes place from this ground state directly to the product state 
(\ref{psi1}) of energy
$E_{0,\ldots,0}$ (\ref{ene1}) with $S^n_p=0$ for all
$n=1,\ldots,N_p$ .
Then the critical $J^c_f$ is defined by $E^{0}_{1,\ldots,1}|_{J_f=J^c_f} =
E_{0,\ldots,0}|_{J_f=J^c_f}$.
According to (\ref{ene2}) we have
$E^{0}_{1,\ldots,1}|_{J_f=J^c_f} =
E^{0}_{1,\ldots,1}|_{J_f=0} + \frac{1}{4}J^c_fN_p \stackrel{!}{=}
-\frac{3}{4}
(J_d+J^c_f)N_p$  which yields
\begin{equation}\label{jc}
J^c_f = -\frac{3}{4} J_d -\frac{1}{N_p} E^{0}_{1,\ldots,1}|_{J_f=0}.
\end{equation}

First we look for an upper bound for $J^c_f$. We consider $J_f \le
J^c_f$. Then the state with $S^n_p=1$ for all
$n=1,\ldots,N_p$ is the ground state and
$E_{0,\ldots,0}$ (\ref{ene1}) sets an upper bound for the
ground-state energy $E^{0}_{1,\ldots,1}$.

\begin{figure}
\begin{center}{\input{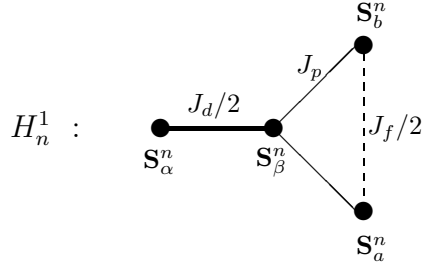}}\end{center}  
\caption{\label{fig3}
Elementary cluster $H_n^1$ for the decomposition $H=\sum
(H^1_{n}+H^2_{n})$ \hspace{0.2cm} (see text).
The cluster $H^2_{n}$ is the mirror immage
of $H^1_{n}$.
}\end{figure}

A lower bound of the ground-state energy  is found (see for instance
\cite{valenti91}) by a simple cluster
decomposition of $H$
\begin{eqnarray}\label{hclust}
H & = & \sum_{n=1}^{N_p} H^1_{n} + H^2_{n} \nonumber \\
H^1_{n} & = & \frac{J_d}{2} {\bf S}^n_{\alpha}{\bf S}^n_{\beta} +
J_p {\bf S}^n_{\beta}({\bf S}^n_{a} + {\bf S}^n_{b}) +
\frac{J_f}{2} {\bf S}^n_{a}{\bf S}^n_{b} \nonumber \\
H^2_{n} & = & \frac{J_f}{2} {\bf S}^n_{a}{\bf S}^n_{b} +
J_p {\bf S}^{n+1}_{\alpha}({\bf S}^n_{a} + {\bf S}^n_{b}) +
\frac{J_d}{2} {\bf S}^{n+1}_{\alpha}{\bf S}^{n+1}_{\beta}
\end{eqnarray}
(see Fig.\ref{fig3}). The lowest energy of $H^1_n$ and $H^2_n$ with $S^n_p=1$ is
\begin{equation}\label{eclust}
E_{n} =
-\frac{J_d}{8} - \frac{J_p}{4} + \frac{J_f}{8} -
\frac{1}{4}\sqrt{J_d^2 -2J_dJ_p + 9J_p^2}.
\end{equation}
The lower bound for
$E^{0}_{1,\ldots,1}$ is $2N_p E_{n}$.
The resulting inequality
$-\frac{3}{4} (J_d+J_f)N_p \ge E^{0}_{1,\ldots,1}
\ge 2N_p E_{n} $
implies the following upper bound for $J^c_f$
\begin{equation}\label{bound1}
J^c_f \le -\frac{J_d}{2} +\frac{J_p}{2} +
\frac{1}{2}\sqrt{J_d^2 -2J_dJ_p + 9J_p^2}
\end{equation}

Next we look for a lower bound for $J^c_f$.
We use Eq. (\ref{jc}) and replace $E^{0}_{1,\ldots,1}({J_f=0})$ by a
variational energy $E_{var}$ of a trial ground state of $H_{p-d}$.
Since $E_{var} \ge
E^{0}_{1,\ldots,1}({J_f=0})$ the lower bound is
\begin{equation}\label{bound2}
J^c_f \ge  -\frac{3}{4} J_d -\frac{1}{N_p} E_{var}.
\end{equation}
For $J_p > 0$ we are able to find a trial state (see section 3)
with $E_{var} \le -N_p \frac{3}{4} J_d $, i.e. $J^c_f \ge 0$ is valid for
finite $J_p$.

\subsection*{\bf
(viii) {\bf Validity of the Marshall-Peierls sign rule in a frustrated spin
system:}}

In the limit $J_f=0$ ($H=H_{p-d}$) the lattice is bipartite and
the Marshall-Peierls sign rule is valid \cite{marsh55,ri94}, i.e. the
phase relations of the ground-state wave function are exactly known.
Though there are several arguments that these sign rule will survive a
finite frustration \cite{kitatani92,retzlaff93,ri94,parkinson95,voigt97}
the validity of the sign rule in nonbipartite frustrated lattice cannot
be shown generally.

Based on statements (vi) and (vii) we argue that for all $J_f < J^c_f$
($J^c_f > 0$) the ground state of the nonfrustrated system $H_{p-d}$
remains.
Hence the considered plaquette-dimer chain is
one example, where the Marshall-Peierls sign rule indeed survives finite
frustration.

\subsection*{\bf
(ix) {\bf Spin gap for large frustration $J_f > J^c_f$:}}

We consider the gap $\Delta$ of the first triplet excitation  versus
the singlet product
ground state  $|\Psi_{0,\ldots,0}\rangle$ (\ref{psi1}).
Since in $|\Psi_{0,\ldots,0}\rangle$  the top and bottom  spins of any
plaquette  are separated from all other spins, the first triplet
excitation is a state $|\Psi_{0,\ldots,0,1,0,\ldots,0}\rangle$
with one triplet for a certain plaquette $i$, i.e. $S^i_p=1$
\begin{equation}\label{psi2}
|\Psi_{0,\ldots,1,0,\ldots,0}\rangle = |\Psi_{-<>-}\rangle
\prod_{n=1}^{N_p-1}|(a_n,b_n)\rangle
\prod_{n=1}^{N_p-2}|(\alpha_n,\beta_n)\rangle \end{equation}
where in $\prod_{n=1}^{N_p-1}$ the plaquette $i$ is excluded and in
$\prod_{n=1}^{N_p-2}$
the left and right neighbouring dimers of plaquette $i$ are excluded.
$|\Psi_{-<>-}\rangle$ represents just the state with $S^i_p=1$
for the excluded plaquette $i$ and the adjacent dimers $i$ and $i+1$.
 The degeneracy of the state (\ref{psi2}) is $3N_p$.

The excitation gap is the energy difference between the ground state
$|\Psi_{0,\ldots,0}\rangle$ (\ref{psi1}) and
$|\Psi_{0,\ldots,0,1,0,\ldots,0}\rangle$ (\ref{psi2})
\begin{equation}\label{gap}
\Delta = E_1 - E_0 = \frac{3}{2}J_d + J_f + E_{-<>-}(J_d,J_p),
\end{equation}
where $E_{-<>-}(J_d,J_p)$
is the energy of the excluded cluster
$-\hspace{-7pt}<\hspace{-2pt}>\hspace{-7pt}-$ for $J_f=0$.
Obviously,
$E_1 - E_0$ is independent of size $N$.
Since
$|\Psi_{-<>-}\rangle$ is a state of only 6 spins there is no problem to
calculate $E_{-<>-}(J_d,J_p)$, i.e. to find the  exact value for $E_1 - E_0$.

\section{Exact diagonalisation versus
perturbation theory }
Using Lanczos algorithm we calculate the ground state and several low-lying
states for chains with periodic boundary conditions of size
$N=8,16,24,32$ (i.e. $N_p=2,4,6,8$ plaquettes).

In the limits of $J_d/J_p \ll 1$ and $J_p/J_d \ll 1$  we calculate
the energies of the singlet
ground state and the first triplet excitation by second order perturbation
theory.  In the limit $J_d=0$ the unperturbed ground state is a
product of the lowest four-spin
plaquette states. In the opposite limit ($J_p=0$)  the ground state
of $H_{p-d}$ (\ref{ham1}) is a
product of dimer singlets and threefold degenerated triplets
of the plaquette spins. The second order perturbation theory in $J_p$
leads to the following effective spin Hamiltonian
\begin{equation}\label{heff}
 H_{eff} = - \frac{3}{4}J_d N_p - \frac{J_p^2}{J_d} N_p+
 \frac{J_p^2}{2J_d} \sum_{n=1}^{N_p} {\bf S}^n_{ab} {\bf S}^{n+1}_{ab}
\end{equation}
where the quantities ${\bf S}^n_{ab}$ are spin $1$ operators, i.e.
$({\bf S}^n_{ab})^2=2$.
We see that in the limit  $J_p/J_d \ll 1$ the dimer-plaquette $S=1/2$ model
maps onto the $S=1$ Haldane chain with an effective
exchange parameter $J_{eff}=J_p^2/2J_d$.

\subsection{Groundstate energy, low-lying excitations, spin gap}

First we consider the ground-state energy $E_0$.
For large frustration $J_f \ge J^c_f$ the explicit expression for $E_0$
is given in Eq. (\ref{ene1}).

For $J_f < J^c_f$ the ground-state energy
obeys Eq. (\ref{ene2}) and it is sufficient to consider the
unfrustrated Hamiltonian $H_{p-d}$.
First we consider the two limits of small dimer exchange
$J_d \ll J_p$ and of small plaquette exchange
$J_p \ll J_d$ .  In the first case the perturbation theory yields
\begin{equation}\label{pert1}
\frac{E_0}{4N_p} = - J_p\left[\frac{1}{2}+
\frac{1}{4} \frac{43}{576}\left(\frac{J_d}{J_p}\right)^2\right].
\end{equation}
The opposite limit is described by the  effective spin $1$ model
(\ref{heff}), i.e.
\begin{equation}
\frac{E_0}{4N_p} = - \frac{3}{16}J_d - \frac{J_p^2}{4J_d} +
 \frac{J_p^2}{8J_d}  \epsilon_H
\end{equation}
where $\epsilon_H = -1.401484038971$ \cite{white93,golinelli94}
is the well-known energy per site of the Haldane chain.

The numerical data for N=16 are drawn in Fig.\ref{fig4}.
The presented energies belong to the corresponding
lowest eigenstate for a given set
of local quantum numbers $S^n_p$, $n=1,\ldots,4$.
(An exception is the state with total spin $S=1$ and
$S^n_p=1$ $n=1,\ldots,4$, which is the first triplet excitation versus
the singlet ground state.) The state with highest energy is just the
product state $|\Psi_{0,\ldots,0}\rangle$ (\ref{psi1}), its energy is
independent of $J_p$ (see Eq. (\ref{ene1})).
All presented energies are degenerated for $J_p=0$. Increasing $J_p$
it follows a quadratic dependence on  $J_p$ for small $J_p$,
and a linear dependence on  $J_p$ for larger $J_p$.
Notice, that the linear dependence on $J_p$ (obtained by perturbation
theory in the limit $J_p \gg J_d$, see (\ref{pert1}))
is well established already
for $J_p \approx J_d$.

In all finite systems ($N=8,16,24,32$) considered in this paper
the first triplet excitation has the same local quantum numbers as the
ground state, i.e. $S^n_p=1$ ($n=1, \ldots ,N_p$). The eigenstates with
singlets $S^n_p=0$ for some $n$ have higher energies and and we find
that the larger the
number of plaquettes with $S^n_p=0$ the higher the energy.

Next we consider the excitation gap $\Delta$
between the singlet ground state and
the first triplet excitation. This triplet excitation is the lowest
excitation at all (see Fig.\ref{fig4}).
The perturbation theory  for large $J_p/J_d$  yields
\be\label{gap1}
\Delta = J_p \big [ 1 - \frac{1}{3} \frac{J_d}{J_p}
- \frac{61}{576} \big(\frac{J_d}{J_p}\big)^2\big].
\ee
This result was already obtained in  \cite{katoh95}.
In the opposite limit $J_d/J_p \gg 1$ we can use the results for the
Haldane chain \cite{white93,golinelli94} and we have
\be\label{gap2}
\Delta = \frac{J^2_p}{2J_d}\Delta_H = 0.41050 \frac{J^2_p}{2J_d}.
\ee
If we include frustration the situation is not changed for $J_f<J^c_f$
except in in a small parameter region in
 the vicinity of the
transition, i.e for $J_f \approx J_f^c$ the first excitation is not a
triplet but a singlet indicating strong frustration effects
\cite{lhuillier}.
For  $J_f>J^c_f$ the gap is exactly known in the whole parameter range
 (see equation (\ref{gap})).

Numerical data are shown in Fig.\ref{fig5}, where $\Delta$ is drawn versus $J_p$.
The linear and quadratic dependences for large and small $J_p$
correspond to the perturbation theory (Eqs. (\ref{gap1}) and
(\ref{gap2}).
In the scale used for Fig.\ref{fig5} the data for $N=24$ and $N=32$ almost
coincide.

In case of the unfrustrated dimer-plaquette chain with identical NN
 bonds $J_p=J_d=1$ the gap is already about $50\%$ larger then the
Haldane gap $\Delta_H$, namely $\Delta_{N=24}=0.60922$,
$\Delta_{N=32}=0.60906$,   $\Delta_{N \rightarrow \infty}=0.6086$.
%
In accordance with 2D models for $CaV_4O_9$ the frustration
may enlarge the gap, in the considered model we have
$\Delta(J_f+x) \ge \Delta(J_f)$ ($x > 0$).

Obviously, though we have a spin $1/2$ chain the $\Delta$ is finite for any
finite $J_p$ which corresponds to the observation that the gapless
spectrum of the Bethe chain is an exceptional case.

\subsection{ Critical $J^c_f$}

The critical point $J^c_f$ is defined in section 2, statement (vii).
This point coincides with the point of
maximal frustration indicated by a
 maximum in the ground-state energy versus $J_f$ precisely at $J_f=J_f^c$.

Upper and lower bounds for  $J^c_f$ are given in Eqs. (\ref{bound1})
and (\ref{bound2}).

For the estimation of the lower bound we consider a variational state of
the form
\begin{equation}\label{psi_var}
|\Psi_{var}\rangle = \prod_{n=1 \atop n \hspace{3pt}
odd}^{N_p-1}|\uparrow_{a_n}\uparrow_{b_n}\rangle
|\downarrow_{a_{n+1}}\downarrow_{b_{n+1}}\rangle
 |\{\alpha_n,\beta_n\}\rangle
|\{\beta_{n+1},\alpha_{n+1}\}\rangle
\end{equation}
where $|\uparrow_{a_n}\uparrow_{b_n}\rangle$
\hspace{4pt}
($|\downarrow_{a_{n+1}}\downarrow_{b_{n+1}}\rangle$)
\hspace{3pt} is a triplet
state of the plaquette spins with z-component $S^{n,z}_{ab} = +1$
($S^{n+1,z}_{ab} = -1$)
and
$|\{\alpha_n,\beta_n\}\rangle |\{\beta_{n+1},\alpha_{n+1}\}\rangle = \\
(1+x^2)^{-1}
\big(|\uparrow_{\alpha_n}\downarrow_{\beta_n}\rangle -
x |\downarrow_{\alpha_n}\uparrow_{\beta_n}\rangle\big)
\big(|\uparrow_{\beta_{n+1}}\downarrow_{\alpha_{n+1}}\rangle -
x |\downarrow_{\beta_{n+1}}\uparrow_{\alpha_{n+1}}\rangle\big) $
is a  variational state which
interpolates between a dimer singlet
state ($x=1$) and a \Ne\ state ($x=0$). The calculation of the optimised $x$
is simple, $x=-2J_p/J_d + \sqrt{1 \hspace{3pt} + \hspace{3pt}4J^2_p/J^2_d}$.
The energy of this state $E_{var}$ entering Eq. (\ref{bound2})
is quite good in the limit of small
$J_p$ and becomes exact for $J_p=0$.

The numerical results are presented in Fig.\ref{fig6}.
While the lower bound
demonstrates that $J^c_f$ is finite for any finite $J_p$
we see the expression (\ref{bound1}) for the upper bound
is close to
the actual value of $J^c_f$ and can serve as an approximative analytic
expression for $J^c_f$.

\subsection{ Pair spin correlation and string order}

At first we consider the limit of large frustrating $J_f > J^c_f$, where
the simple product state $|\Psi_{0,\ldots,0}\rangle$
(\ref{psi1}) is the ground state. Then all spin-spin correlations are
zero except $\langle {\bf S}^n_{\alpha}{\bf S}^n_{\beta}\rangle$ and
$\langle {\bf S}^n_{a}{\bf S}^n_{b}\rangle$ which take there extreme
value $-3/4$.

In what follows we discuss the more interesting case $J_f < J^c_f$, i.e.
the ground state is that of the unfrustrated $H_{p-d}$.
Numerical results for $N=24$ and $N=32$ are shown in Figs.
\ref{fig7},\ref{fig8},\ref{fig9},\ref{fig10}.
To get a general impression on the distance dependence of the correlations
we present in Fig.\ref{fig7} a histogram of the pair correlation versus
separation for three
values of $J_p/J_d$.
The short range correlations
$\langle {\bf S}^n_{\alpha}{\bf S}^n_{\beta}\rangle$ (NN
dimer spins), $\langle {\bf S}^n_{\beta}{\bf S}^n_{a(b)}\rangle$
(NN plaquette-dimer spins) and
$\langle {\bf S}^n_{a(b)}{\bf S}^{n+1}_{a(b)}\rangle$
(plaquette spins of two neighbouring plaquettes) versus $J_p/J_d$ are
drawn in Fig.\ref{fig8} and
the  spin correlations for large separations, namely
 $\langle {\bf S}^n_{a(b)}{\bf S}^{n+3}_{\beta}\rangle$
(dimer spin -- plaquette spin),
$\langle {\bf S}^n_{a(b)}{\bf S}^{n+3}_{a(b)}\rangle$
(plaquette spin -- plaquette spin)  and
 $\langle {\bf S}^n_{\alpha}{\bf S}^{n+3}_{\beta}\rangle$
(dimer spin -- dimer spin) are given in
Fig.\ref{fig9}.

In the dimer limit ($J_p \ll J_d$) the dimer and plaquette spins are
decoupled, i.e. $\langle {\bf S}^n_{\alpha(\beta)}{\bf
S}^m_{a(b)}\rangle =0 $. Otherwise, the NN
dimer correlation $\langle {\bf S}^n_{\alpha}{\bf S}^n_{\beta}\rangle$
takes its extreme value $-3/4$ while for spins belonging to different
dimers
$\langle {\bf S}^n_{\alpha(\beta)}{\bf S}^m_{\alpha(\beta)}\rangle$
($n \ne m$)
goes to zero, too.
However,
though the  dimer and plaquette spins and the non-neighbouring dimer
spins are not correlated there is
a well-pronounced correlation between more
distant plaquette spins with several dimer spins in
between. This is a typical quantum effect; a classical spin chain
with NN exchange
would be split in separated pieces  at that point where the NN correlations
are zero.
The correlation between spins of different plaquettes is described by
the effective Haldane chain (\ref{heff}), the numerical results indicate
that this effective Hamiltonian describes the chain well until $J_p
\approx 0.1 \ldots 0.15J_d$.
For example for $J_p=0.1J_d$ the plaquette--plaquette
correlation $\langle {\bf S}^n_{a(b)}{\bf S}^{m}_{a(b)}\rangle$
differs from the corresponding Haldane correlation $\frac{1}{4}
\langle {\bf S}_{n}{\bf S}_{m}\rangle$ (indicated by crosses $+$ in
Fig.\ref{fig7}) by less then $3\% $.

In the plaquette limit ($J_p \gg J_d$) the ground state becomes a simple
product state of the lowest four-spin
plaquette states. Hence, for $J_p \rightarrow \infty$ we have
$\langle {\bf S}^n_{\alpha}{\bf S}^m_{\beta}\rangle \rightarrow 0$,
$\langle {\bf S}^n_{a(b)}{\bf S}^{l}_{a(b)}\rangle \rightarrow 0$
($n \ne l$) and
$\langle {\bf S}^n_{\beta}{\bf S}^n_{a(b)}\rangle \rightarrow -0.5$.
Already for $J_p = J_d$ the pair correlation drops down very rapidly
(cf. Fig.\ref{fig7}) and besides
the correlation along the $J_p$ bond,
$\langle {\bf S}^n_{\beta}{\bf S}^n_{a(b)}\rangle$,
only extremely short-ranged correlations are present.

There is comparably small region around $J_p
\sim 0.3 \dots 0.4 J_d$ where we have a balance between $J_p$ and $J_d$
and all correlations are well-pronounced.
Since we have a gap for all $J_p > 0$ we argue that all correlations
show exponential decay but with different correlation lengths $\xi_{dd}$ for
the dimer--dimer, $\xi_{dp}$ for the dimer--plaquette  and
$\xi_{pp}$ for the plaquette--plaquette
correlations.
The obtained results suggest that
$\xi_{pp}$ is
is quite large for $J_p/J_d \ll 1$
($\xi_{pp} = \xi_{Haldane} \approx 6.03$
\cite{white93,golinelli94,scholl} for $J_p/J_d \rightarrow 0$),
With increasing $J_p$ there is  a continuous decrease of $\xi_{pp}$
up to $\xi_{pp} \rightarrow 0$ for $J_p \rightarrow \infty$.
Otherwise, $\xi_{dd}$ and $\xi_{pd}$ are extremely small for
$J_p/J_d \ll 1$ and $J_p/J_d \gg 1$ but show a maximum for
$J_p/J_d \sim 0.35 $ ($\xi_{dd}$) and $J_p/J_d \sim 0.3 $ ($\xi_{pd}$) .

Finally we discuss the
string order parameter describing possible hidden order in spin $1$
chains \cite{white93,golinelli94,mikeska,scholl}.
This order parameter is defined as \\
${\cal O}^z_{\pi}(i,j)=\big\langle S^z_i\big (\exp \sum_{k=i+1}^{j}i\pi
S^z_k \big)S^z_j\big\rangle$ where the $S^z_i$ are spin one objects.
For the Haldane spin $1$ chain we have
${\cal O}^z_{\pi} = lim_{|i-j|\rightarrow\infty}
{\cal O}^z_{\pi}(i,j)=0.374325096$  and the value ${\cal
O}^z_{\pi}(1,4)$ for the third neighbour is already close to
${\cal O}^z_{\pi}$ \cite{white93}.
For the considered dimer plaquette chain we write
\be\label{string}
{\cal O}^z_{\pi}(n,m)=\big\langle S^{n,z}_{ab}
\big (\exp \sum_{k=n+1}^{m}i\pi
S^{k,z}_{ab} \big)S^{m,z}_{ab}\big\rangle
\ee
 with $S^{m,z}_{ab}$ defined in
(\ref{com}).
The results are shown in Fig.\ref{fig10}.
In agrrement with pair correlation we observe a Haldane like behaviour
until  $J_p \approx 0.1 \ldots 0.15J_d$ which is followed by
region  $0.15 J_d \stackrel{<}{\sim} J_p \stackrel{<}{\sim} 0.6J_d$  where a
crossover from the Haldane behaviour to the product state with
vanishing  pair correlations and vanishing string order takes place.

\section{Conclusions}

We have calculated the ground-state properties and low-lying excitations
for a $S=1/2$ chain
with alternating dimers and plaquettes (see Eqs. (\ref{ham1}) and
(\ref{ham2})
and Fig.\ref{fig1}).
This model
is in some sense the 1D counterpart of the $1/5$ depleted
square lattice Heisenberg model for $CaV_4O_9$.

While  the classical
ground state  of unfrustrated model $H_{p-d}$ is the \Ne\ state there is a
quantum competition between local singlet formation on the dimers or on
plaquettes for $S=1/2$.
Besides of exact diagonalisation and perturbation theory results
we have given several general
and rigorous statements.

The main results can be summarized as follows.
The ground-state properties and a
class of excitations of $H_{p-d}$ can be mapped on a mixed
spin$-1/2-$spin$-1$ chain with two dimer $S=1/2$ spins and
one effective $S=1$ plaquette spin in the unit cell.
In the limit of small plaquette bonds $J_p \ll J_d$ the ground-state
correlations of the effective $S=1$ plaquette spins
can be described by a Haldane chain.
Increasing the ratio $J_p/J_d$ a crossover takes place from the
effective Haldane chain to a ground state described by a product of
plaquette singlet states.
The pair correlations are characterized by three different correlation
lengths for dimer-dimer, dimer-plaquette and plaquette-plaquette
correlations. In the limit $J_p \ll J_d$
the correlations between plaquette and dimer spins as well as
between non-neighbouring dimer spins vanish, but surprisingly the
correlation between plaquette spins are well pronounced. (Note that this
is a purely quantum effect and has no classical analogue).
In the opposite limit $J_p \gg J_d$
all correlation lengths are extremly short ranged.

Though the dimer-plaquette chain $H_{p-d}$ is a $S=1/2$ model the first
triplet excitation is separated by a gap for all parameter values
except $J_p=0$. This is consistent with the observation
that the gapless ground state of the
Bethe chain is quite unstable against the addition of relevant
operators to create a gap in the excitation spectrum (see for instance
the $S=1/2$
chain with alternating  NN bonds \cite{lc_CuGe}).

Frustration can be introduced in the model in a simple way by adding an
antiferromagnetic interaction of strength $J_f$  between the top and the
bottom spin of a plaquette (see (\ref{ham2}) and Fig.\ref{fig1}).
In the frustrated model we find a first order quantum phase
transition at a finite critical frustration $J^c_f$
between the ground-state phase described above and a
completely dimerized phase, which is similar to a recently described
first-order transition in antiferromagnetic $S=1/2$ coupled chains
\cite{xiang95,nigge97}. Close to the transition the first excitation over
the ground state is not a triplet  but a singlet, which is a signature of
strong frustration \cite{lhuillier}.
The considered model is one example for the rigorous validity of the
Marshall-Peierls sign rule in a frustrated  antiferromagnet.

As mentioned in section 2 we will briefly point out some important 
differences between the dimer-plaquette chain discussed in this paper 
and the spin 1/2 diamond chain considered in
\cite{takano96,nigge97}. 
In the limit of 
small frustration the diamond chain corresponds to a quantum 
ferrimagnet. The ground state of this  ferrimagnet has macroscopic 
total spin $S=N/6$, is
long-range ordered and the spectrum is gapless \cite{ferri1,ferri2,ferri3}. 
On the other hand, the 
dimer-plaquette chain has a singlet ground state without long-range order
and has a gap for all
parameter values considered here.
However,  the common property of both models consists in the 
product singlet state for
large frustration.

Finally we mention that a straightforward extension of the model is
obtained by adding further plaquette spins ${\bf S}^n_c$, ${\bf S}^n_d$,
$\ldots$. The ground-state properties of this extended model could be
mapped on a corresponding mixed
spin-$1/2$-spin-$p/2$ chain ($p$ is the number of spins in a plaquette $n$)
with two dimer $S=1/2$ spins and
one effective $S=p/2$ plaquette spin in the unit cell.

\subsection*{\bf Acknowledgments}

This work was supported by the DFG (Ri 615/1-2) and Bulgarian
Science foundation, Grant F412/94. The authors are
indebted to U.Schollw\"ock for fruitful discussions and to P.Tomczak
for reading the manuscript.

\vspace{10mm}


\begin{figure}\caption{\label{fig4}
Energy eigenvalues  versus $J_p/J_d$ for
the unfrustrated model $H_{p-d}$ with $N=16$
sites. The four numbers in brackets give the local quantum numbers
$S^n_p$, $n=1,\ldots,4$; $S$ is the quantum number of the total spin.
}\end{figure}

\begin{figure}\caption{\label{fig5}
Excitation gap between the singlet ground state and the lowest triplet
excitation. The dashed lines and the squares
correspond to the unfrustrated case. The solid line corresponds
to the frustrated
case with $J_f>J^c_f$, where the gap is independent of $N$.
For $J_f<J^c_f$ tha gap of the frustrated and the unfrustrated model
coincide. The critical point for $N=24$ is at  $J_p = 0.534 J_d$.
}\end{figure}

\begin{figure}\caption{\label{fig6}
Critical frustration $J^c_f$ versus $J_p$ for N=24 (crosses) and N=16
(solid line) and upper and lower bounds (see text). Above the critical
line the ground state of $H$ is the the fully dimerized
product state (\ref{psi1}) and below the critical line the
ground state of the total Hamiltonian
$H$ coincides with that one of $H_{p-d}$ (Eq. (\ref{ham1})).
}\end{figure}

\begin{figure}\caption{\label{fig7}
Spin pair correlations
$\langle {\bf S}(0){\bf S}(j)\rangle$
versus separation of the unfrustrated chain
$H_{p-d}$ of length $N=32$ sites for three values of $J_p/J_d$.
For the spin-spin separation the number of NN steps from the
reference spin at $0$ to
spin at $j$ is taken.
Left (P) side of the Figures -- reference spin ${\bf S}(0)$ is a
plaquette spin ${\bf S}^n_{a(b)}$,
i.e. the correlations $j=1,2,4,5,7,8,10,11$ are
plaquette-dimer correlations, the correlations $j=3,6,9,12$ are
plaquette-plaquette correlations (cf. Fig.\ref{fig1}).
Right (D) side of the Figures -- reference spin ${\bf S}(0)$ is a
dimer spin ${\bf S}^n_{\alpha}$,
i.e. the correlations $j=1,3,4,6,7,9,10,12$ are
dimer-dimer correlations, the correlations $j=2,5,8,11$ are
dimer-plaquette correlations (cf. Fig.\ref{fig1}).
The crosses indicate the correlations of the  corresponding Haldane
chain of length $N=8$.
}\end{figure}

\begin{figure}\caption{\label{fig8}
Short range spin correlations versus $J_p$ for the unfrustrated chain
$H_{p-d}$ and $N=24$ sites.
 $\langle {\bf S}^n_{\beta}{\bf S}^n_{a(b)}\rangle$
$-$ NN dimer spin--plaquette spin,
 $\langle {\bf S}^n_{a}{\bf S}^{n+1}_{a}\rangle$
$-$ plaquette spin -- plaquette spin of neighbouring plaquettes,
$\langle {\bf S}^n_{\alpha}{\bf S}^n_{\beta}\rangle$
 $-$ NN dimer spin -- dimer spin
}\end{figure}

\begin{figure}\caption{\label{fig9}
Spin correlations for largest spin separations versus $J_p$
for the unfrustrated chain
$H_{p-d}$ and $N=24$.
 $\langle {\bf S}^n_{a}{\bf S}^{n+3}_{\beta}\rangle$
 $-$  dimer spin -- plaquette spin,
 $\langle {\bf S}^n_{a}{\bf S}^{n+3}_{a}\rangle$
 $-$  plaquette spin -- plaquette spin,
 $\langle {\bf S}^n_{\alpha}{\bf S}^{n+3}_{\beta}\rangle$
$-$   dimer spin -- dimer spin
}\end{figure}

\begin{figure}\caption{\label{fig10}
String order
${\cal O}^z_{\pi}(n,m)$ (see Eq. (\ref{string}) for
$(n,m)=(1,2)$, $(1,3)$, $(1,4)$ versus $J_p/J_d$ for
the unfrustrated chain
$H_{p-d}$ of length $N=24$ and $N=32$.
}\end{figure}


\newpage\thispagestyle{empty}
\begin{center} {\epsfig{file=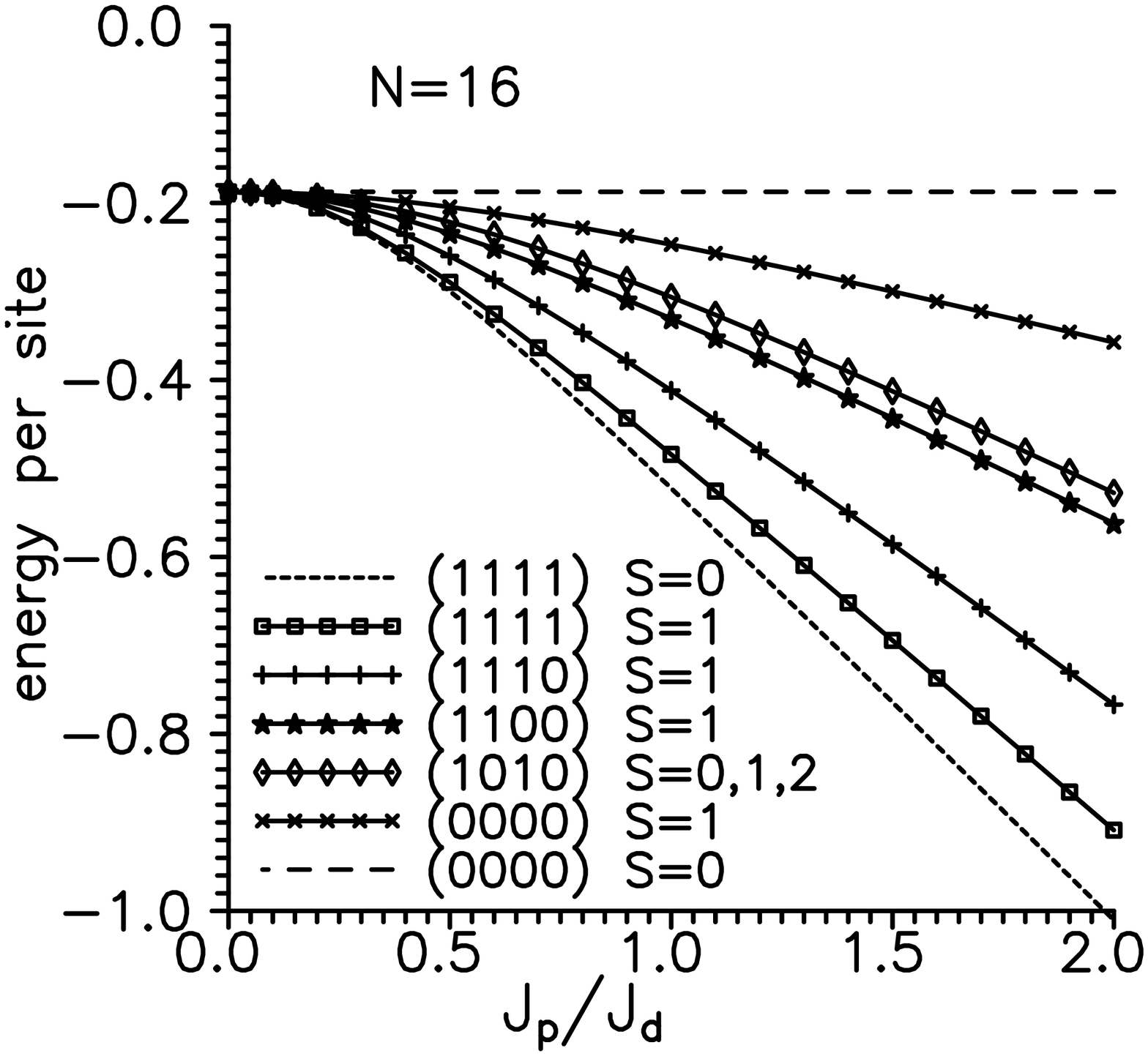,scale=0.6}} \end{center}
\vfill
Fig.4 J.Richter et. al.

\newpage\thispagestyle{empty}
\begin{center} {\epsfig{file=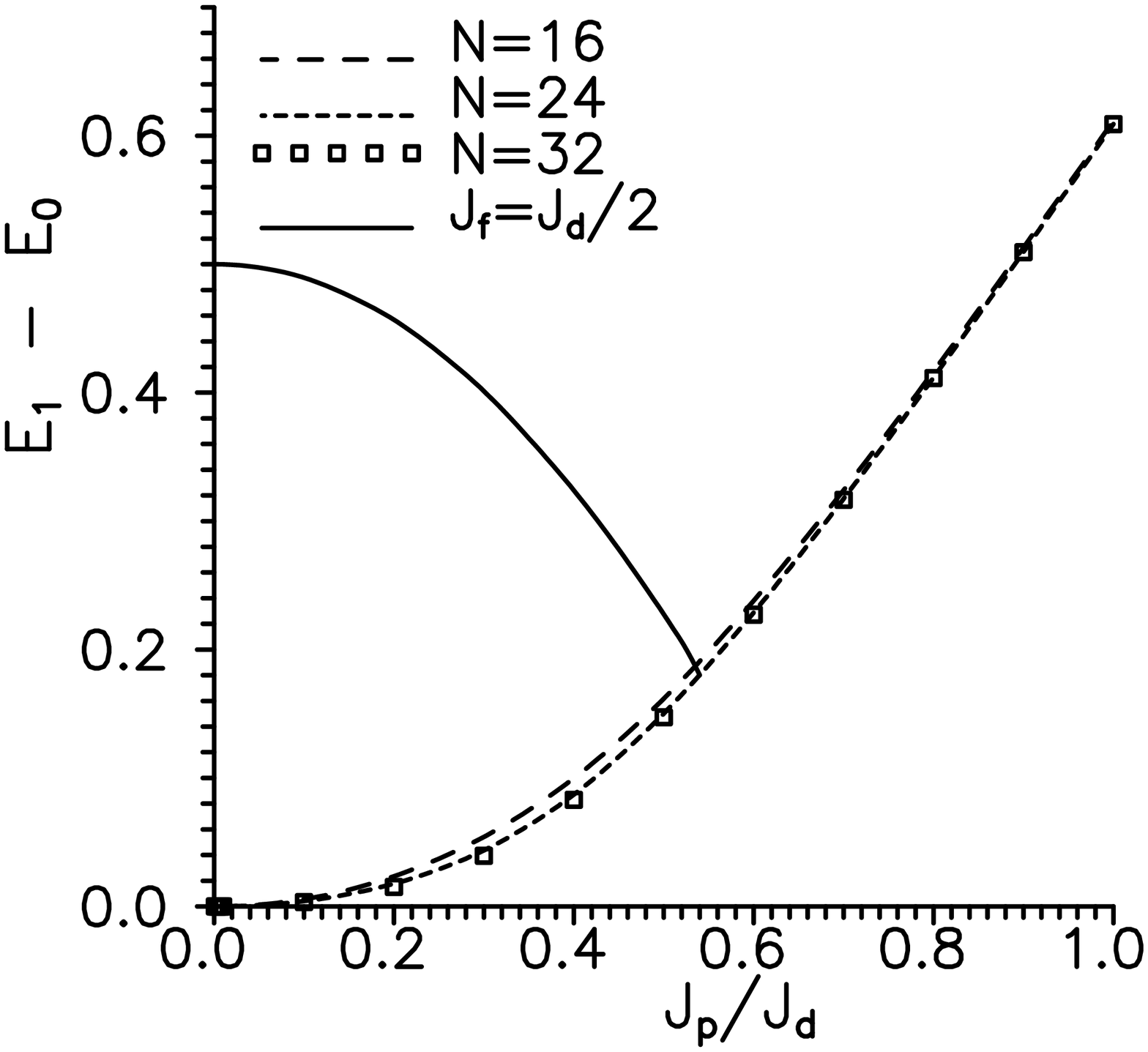,scale=0.6}} \end{center}
\vfill
Fig.5 J.Richter et. al.

\newpage\thispagestyle{empty}
\begin{center} {\epsfig{file=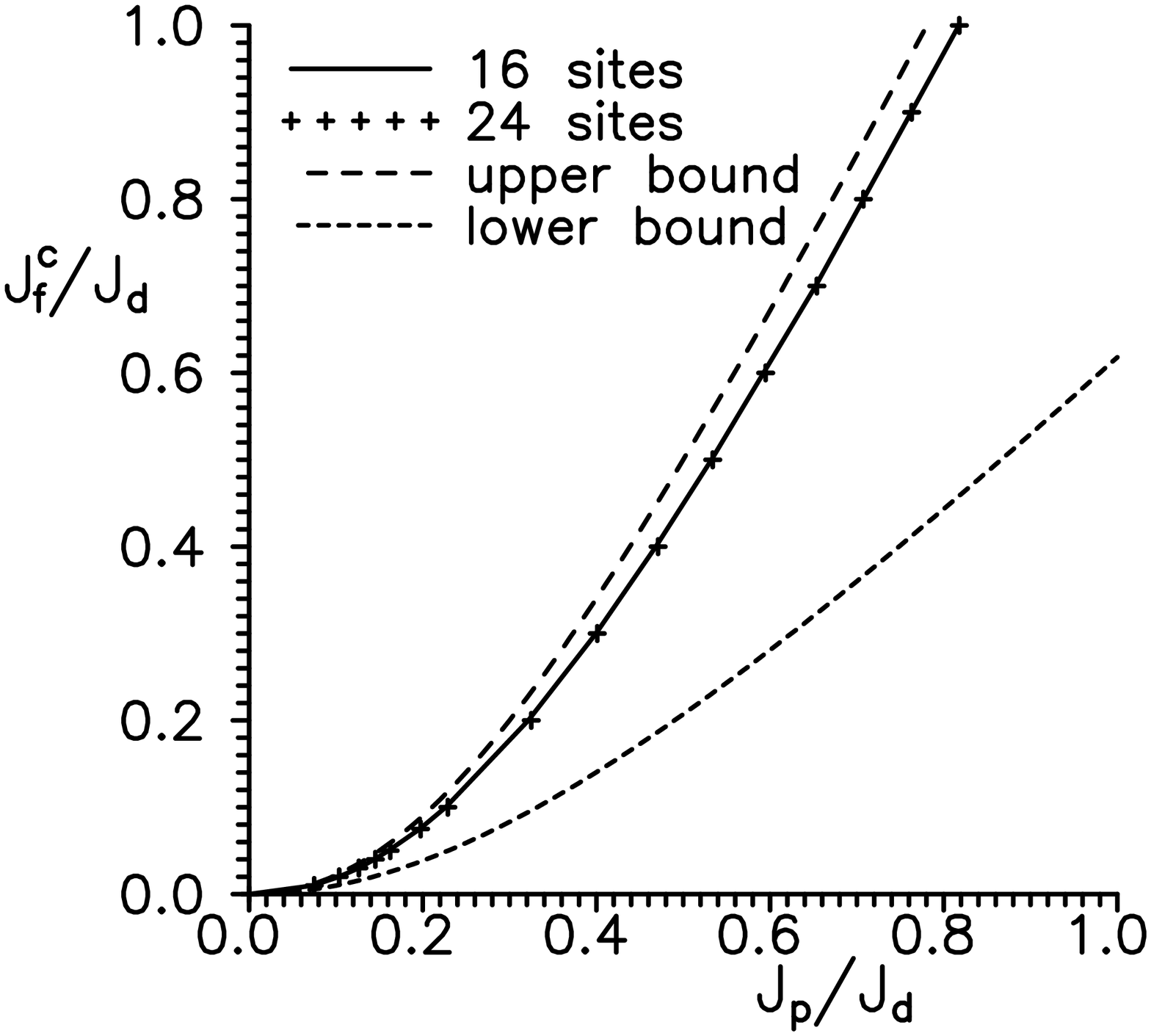,scale=0.6}} \end{center}
\vfill
Fig.6 J.Richter et. al.

\newpage\thispagestyle{empty}
\begin{center} {\epsfig{file=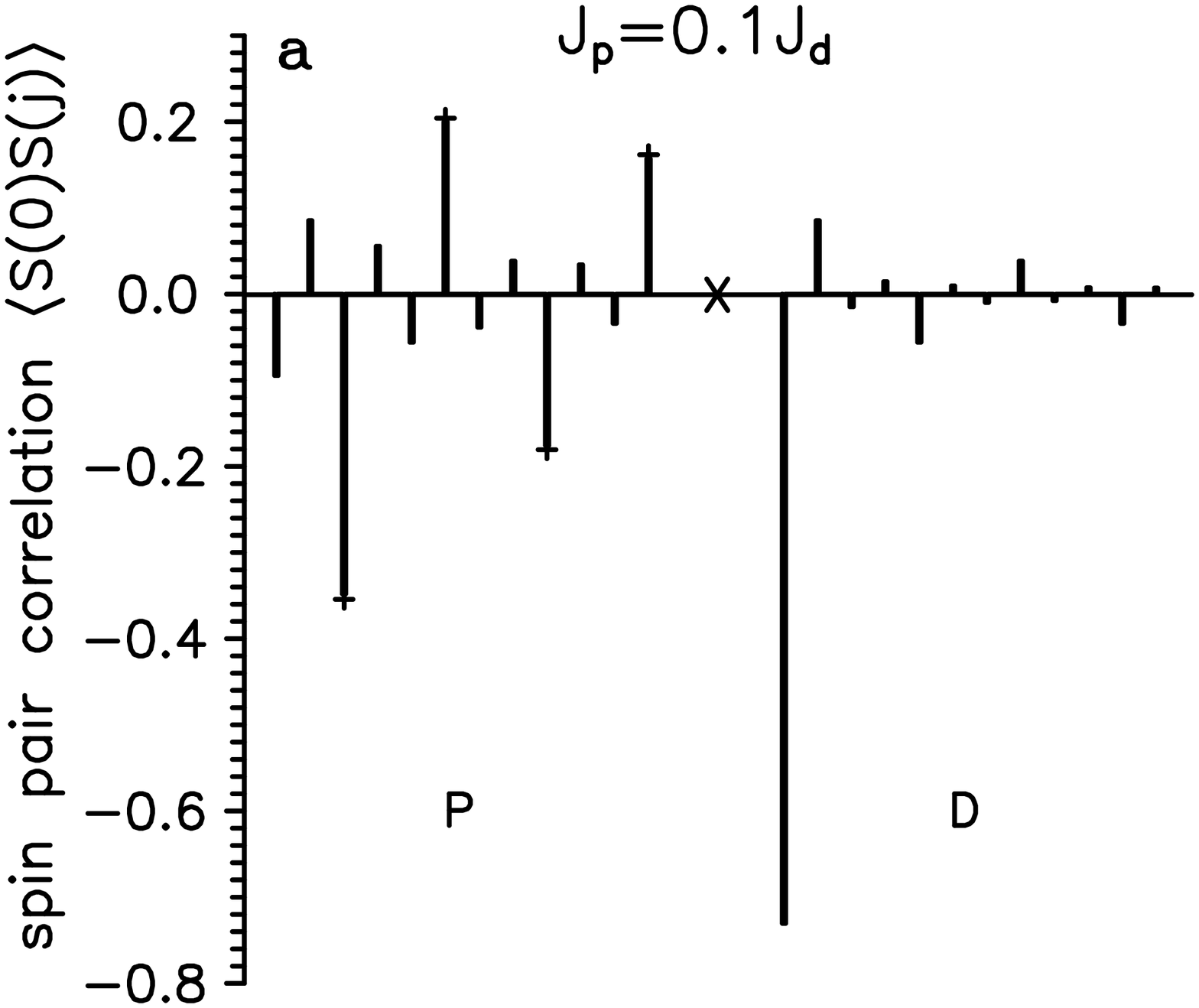,scale=0.6}} \end{center}
\vfill
Fig.7a J.Richter et. al.

\newpage\thispagestyle{empty}
\begin{center} {\epsfig{file=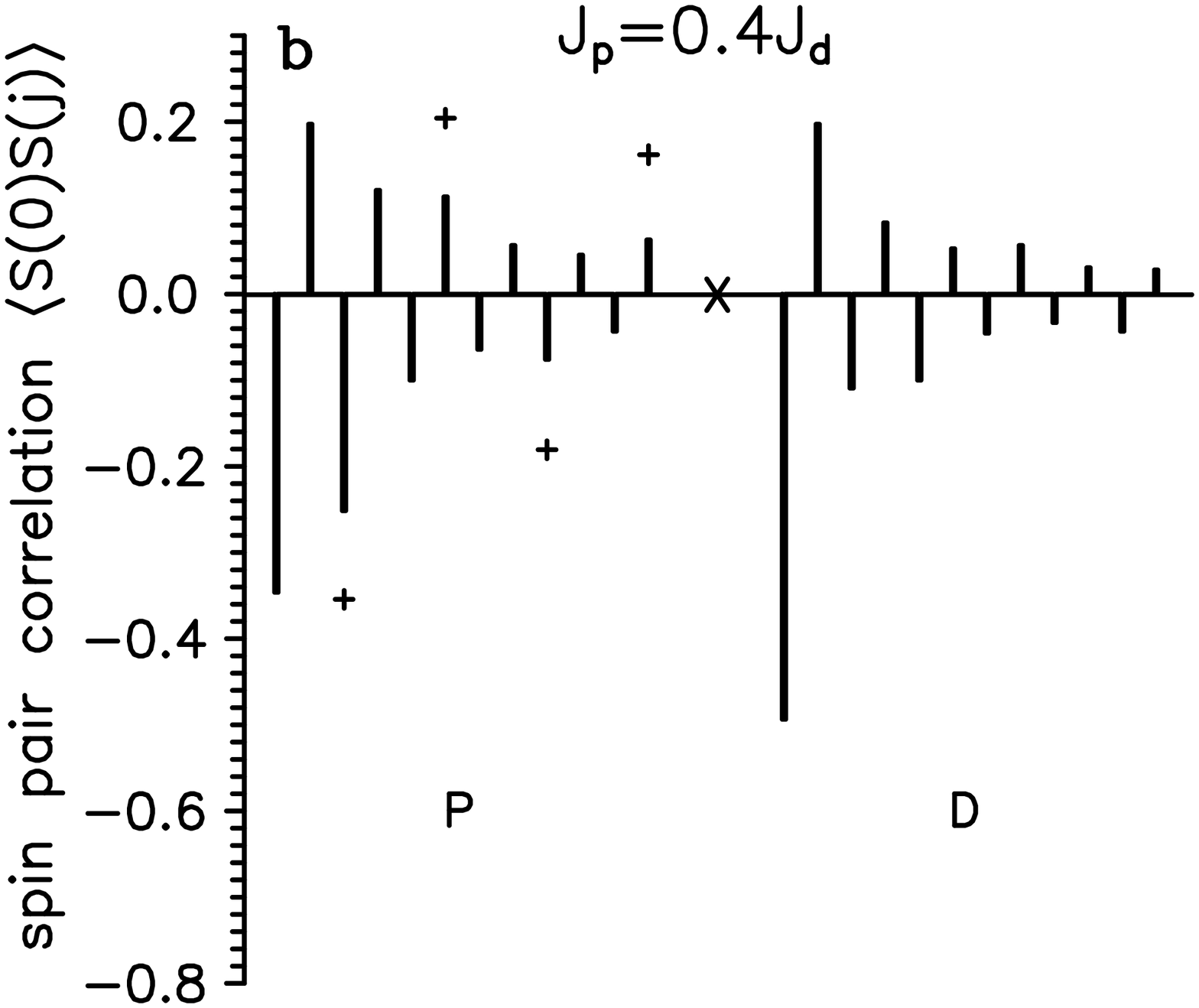,scale=0.6}} \end{center}
\vfill
Fig.7b J.Richter et. al.

\newpage\thispagestyle{empty}
\begin{center} {\epsfig{file=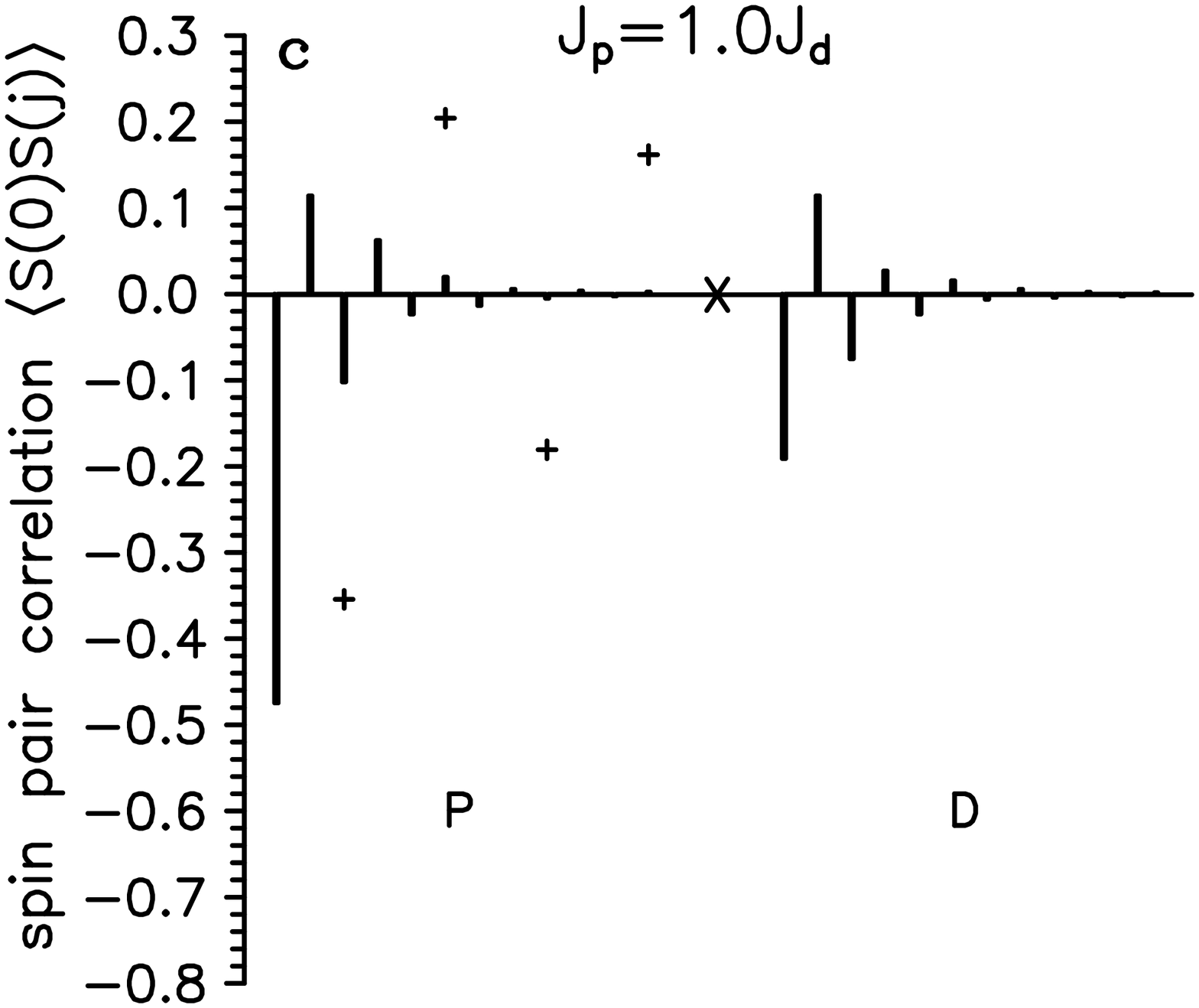,scale=0.6}} \end{center}
\vfill
Fig.7c J.Richter et. al.

\newpage\thispagestyle{empty}
\begin{center} {\epsfig{file=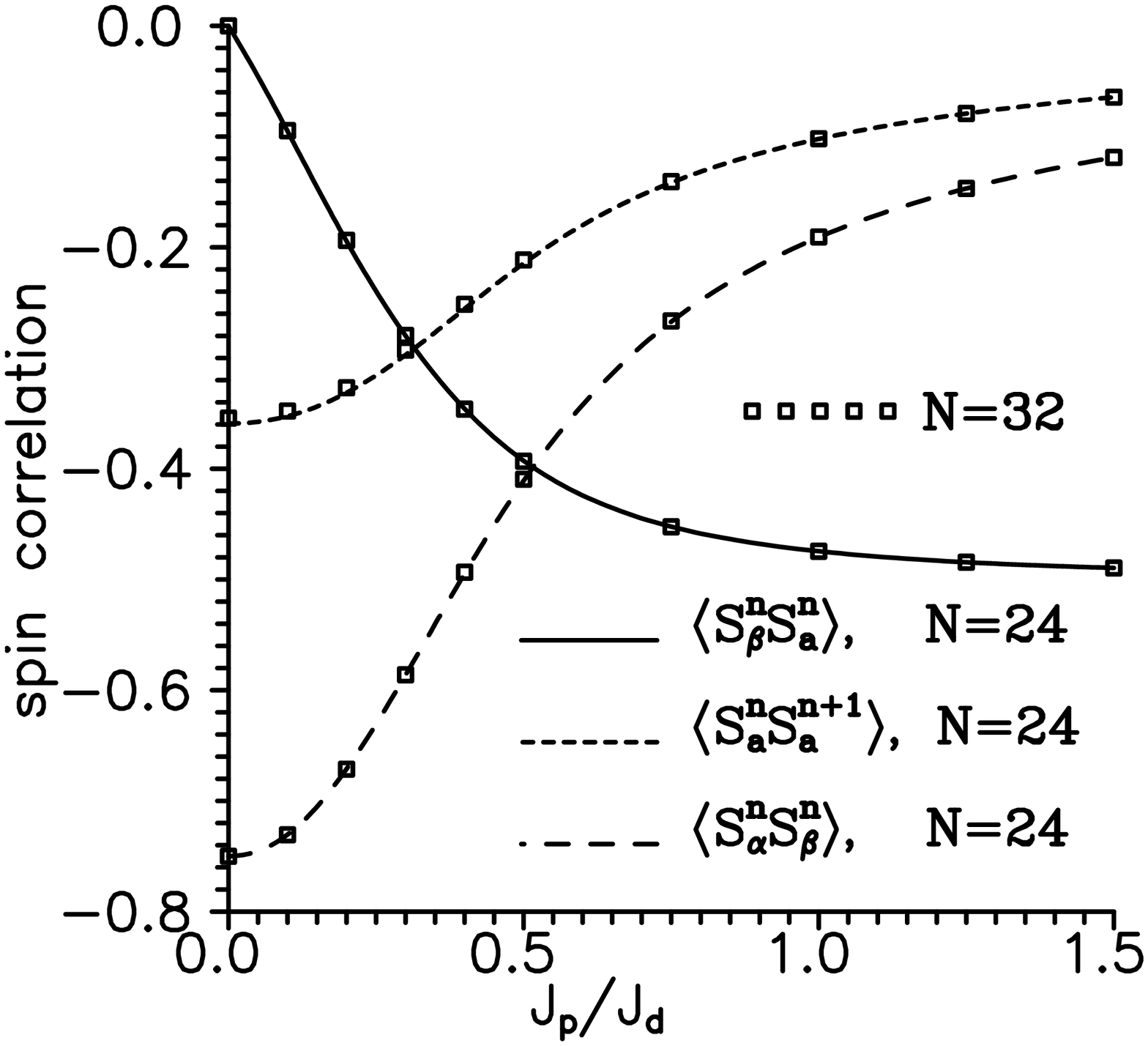,scale=0.6}} \end{center}
\vfill
Fig.8 J.Richter et. al.

\newpage\thispagestyle{empty}
\begin{center} {\epsfig{file=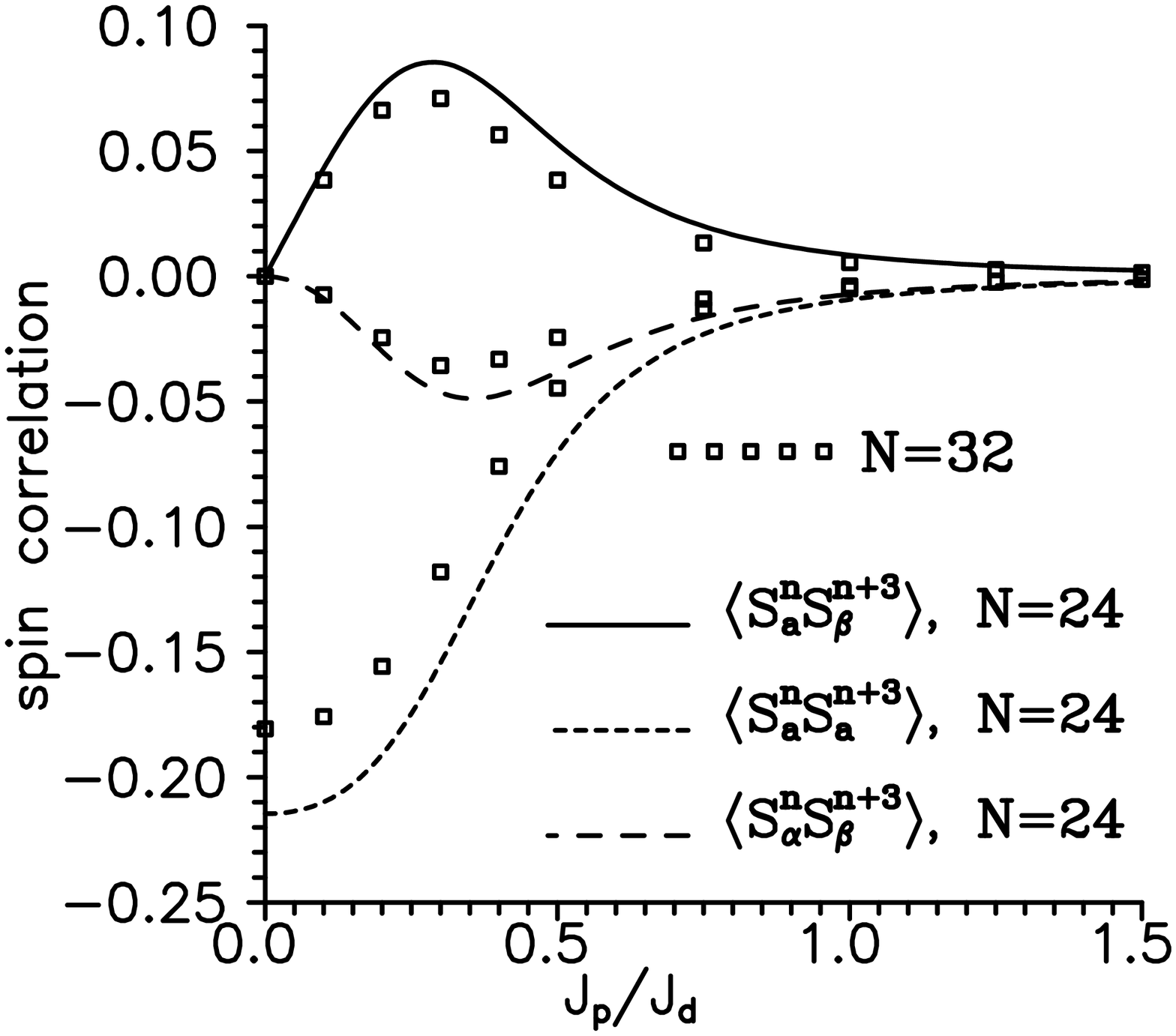,scale=0.6}} \end{center}
\vfill
Fig.9 J.Richter et. al.

\newpage\thispagestyle{empty}
\begin{center} {\epsfig{file=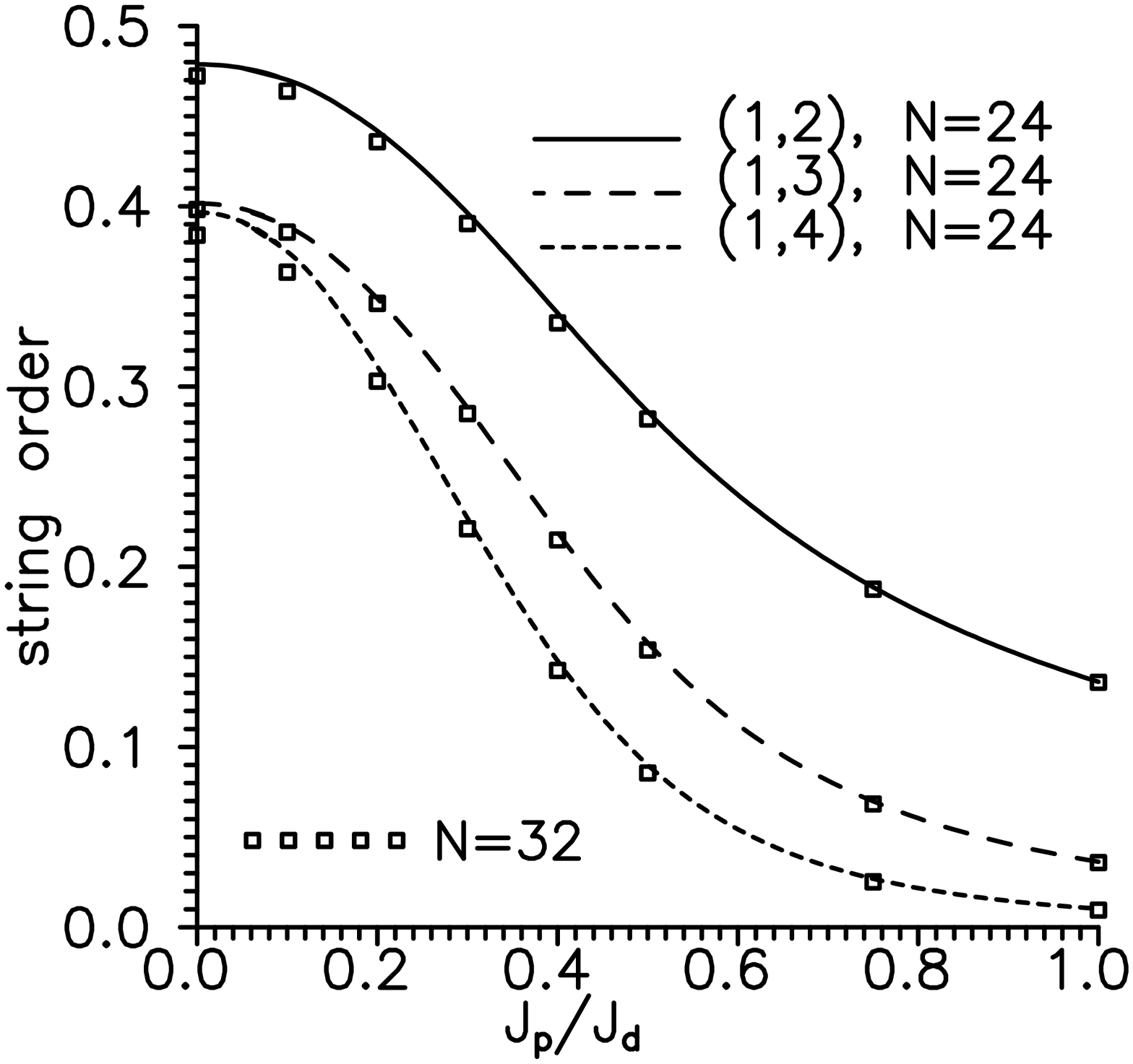,scale=0.6}} \end{center}
\vfill
Fig.10 J.Richter et. al.

\end{document}